\documentclass[english,figures]{epl}
\usepackage{times}
\usepackage[T1]{fontenc}
\usepackage[latin1]{inputenc}
\usepackage{graphicx}
\usepackage{amssymb}

\makeatletter

\providecommand{\boldsymbol}[1]{\mbox{\boldmath $#1$}}

\usepackage{babel}
\makeatother

\title{Analytical study of tunneling times in flat histogram Monte Carlo}
\shorttitle{Analytical study of tunneling times in flat histogram...}

\author{Miguel D. Costa\inst{1,2} \and J. Viana Lopes\inst{1,3}\thanks{E.mail: \email{jlopes@fc.up.pt}} \and J.M.B. Lopes dos Santos\inst{1} }
\shortauthor{Miguel D. Costa \etal}

\institute{
\inst{1} Centro de Física do Porto, Departamento de Física, Faculdade de Ciências, Universidade do Porto, 4169-007 Porto Portugal\\
\inst{2} Escola Superior de Tecnologia e Gestão, Instituto Politécnico de
Viana do Castelo, Viana do Castelo Portugal\\
\inst{3} Departamento de Física, Instituto Superior de Engenharia, Instituto
Politécnico do Porto, Porto Portugal
}


\begin{document}

\maketitle

\begin{abstract}
We present a model for the dynamics in energy space of multicanonical
simulation methods that lends itself to a rather complete analytic
characterization. The dynamics is completely determined by the density
of states. In the $\pm J$ 2D spin glass the transitions between the
ground state level and the first excited one control the long time
dynamics. We are able to calculate the distribution of tunneling times
and relate it to the equilibration time of a starting probability
distribution. In this model, and possibly in any model in which entering
and exiting regions with low density of states are the slowest processes
in the simulations, tunneling time can be much larger (by a factor
of $O(N)$) than the equilibration time of the probability distribution.
We find that these features also hold for the energy projection of
single spin flip dynamics.
\end{abstract}

The simulation of models with complicated energy landscapes, such
as spin glasses or models of protein folding, has always proved impractical
for conventional canonical Monte Carlo methods \cite{NewmanBarkema}.

In a Monte Carlo simulation of a system of $N$ degrees of freedom,
at fixed temperature $T$, energy fluctuations about a mean energy
$E(T)$ are of order $k_{B}T\sqrt{Nc_{v}}$. As a result, only states
in a range of energies $\langle E\rangle-\delta E<E<\langle E\rangle+\delta E$
with $\delta E\sim k_{B}T\sqrt{Nc_{v}}$ are accessible in the simulation.
In systems with complex energy landscapes, phase space in this narrow
energy range breaks up into many regions, connected only by states
requiring energy fluctuations $\Delta E\gg\delta E$; tunneling times
between these regions become too long to retain any hope of achieving
the asymptotic distribution in a reasonable simulation time. 

To overcome this problem, one would like to broaden the energy range
of the states sampled in a simulation, forsaking the canonical ensemble
at a fixed temperature. A variety of methods have been proposed to
implement this idea, such as entropic sampling \cite{LEE93}, multicanonical
Monte Carlo \cite{BN92,BN91}, simulated and parallel tempering\cite{MP92,HN96,LMS+92},
Wang-Landau sampling \cite{WL01,WL01b,TWA03}, broad histogram and
transition matrix methods \cite{Oliveira96,WS02,WTS99,THT04}. In
some cases, like multicanonical Monte Carlo and Wang-Landau sampling,
the aim is to sample all energy levels of the spectrum with equal
probability, producing flat histograms in energy space. While this
allows overcoming large energy barriers in systems with rugged energy
landscapes, critical slowing down can remain a problem \cite{DTW+04,THT04}.

Given the practical impossibility of measuring equilibration time
of an initial distribution, the performance of these methods is generally
assessed by measuring the so called tunneling time \cite{DTW+04,LCLdS,BERGCElik92},
the time required to cross the entire energy spectrum, from ground
states, to states with the energy of maximum density of states, and
back, or from ground state to anti-ground states, states of maximum
energy.

In this letter we focus on the dynamics of this random walk in energy
space. We propose a modification of the directed network of non-zero
transition probabilities which underlies the Markov chain of a Monte
Carlo simulation, which, while retaining the long time behaviour of
the energy projection of conventional simulations, lends itself to
a very complete analytic description. Our three most important results
concern the $\pm J$ 2D spin glass: (a) the transitions between ground
state level and first excited one completely control the long time
dynamics of simulations in this model; (b) the equilibration time
of a simulation can be considerably shorter than tunneling time; (c)
these conclusions also hold for the case of conventional single spin
flip (SSF) dynamics. 

To calculate thermodynamic averages in any system, we need only specify
the set of points of phase space and the corresponding probabilities.
To solve this problem using Monte Carlo, we impose on this structure
a directed network of non-zero transition rates which defines the
Markov chain used to generate the asymptotic distribution. The traditional
SSF connects each point of phase space to $N$ first neighbours. The
SSF algorithm is, usually, local in energy, i.e., energy differences
between connected phase space points being of $O(1)$, independent
of system size. It has been claimed that differences in the scaling
behaviour of the $\pm J$ 2D spin glass model and a fully frustrated
model, both with extensive ground-state entropy, reside in the restrictions
imposed by the network defined by SSF dynamics. Still, the scaling
laws of relevant time scales with system size found with our dynamics
\cite{LCLdS} are also similar to those obtained with SSF dynamics
\cite{DTW+04}.

The simplest choice that circumvents the problem of local minima is
to construct a Markov chain that connects each point of phase space
to all the remaining points that have the same or adjacent energies.
In this way, we preserve the locality in energy space of the standard
SSF network and avoid the need for long paths to connect states that
are adjacent in energy. 

The procedure in best explained in reference to a specific model like
the $\pm J$ 2D Ising spin glass. The corresponding energy levels,
$E_{i}$, and degeneracies, $N(E_{i})$, can be calculated exactly
using the program of Saul and Kardar \cite{SK94}. Non-zero transition
rates connect states with energy $E_{i}$ to all states with energy
$E_{i}$ or $E_{i\pm1}$. The simplest choice of transition rates
is to propose the final state uniformly from the set of allowed states,
and accept the proposed state with a probability which ensures that
the asymptotic probability of each state equals $1/N(E)$. Whereas
in a SSF simulation most transition probabilities to states of the
same or nearby energy are zero, in our model they are all non zero,
with values adjusted to ensure the same equilibrium distribution.
This procedure is akin to an averaging of transition rates. While
this could lead to significant changes in the dynamics, we nevertheless
find that the most relevant features of the long time SSF dynamics
in energy space still hold in our model. The advantage of the current
model is that the projection of the multicanonical Markov chain in
energy space remains a Markov process, allowing us the use of a set
of analytic tools to study the dynamics. 

We denote by $p_{\alpha}(t)$ the probability of being in a state
$\alpha$; $N_{i}$ is the degeneracy of energy level $E_{i}$ and
$M_{i}\equiv N_{i-1}+N_{i}+N_{i+1}-1$ is the number of states accessible
in a direct transition from any state of energy $E_{i}$ (the choice
of tentative final states excludes the current state). In equilibrium,
the flat energy histogram condition requires, for a state $\alpha$
of energy $E_{i}$, \begin{equation}
p_{\alpha}=\frac{1}{N(E_{\alpha})}\equiv\frac{1}{N_{i}}\label{eq:peq_flat}\end{equation}
For reasons of efficiency, we exclude the negative temperature region:
for energies greater than $E_{m}$, the energy at the maximum of density
of states, we replace $N(E)$ by $N(E_{m})$ in eq.~\ref{eq:peq_flat}.
The transition rate from a state $\alpha$ to an accessible state
$\beta$, $\omega_{\beta\alpha}$, satisfies detailed balance, with
the equilibrium distribution given in eq.~\ref{eq:peq_flat}; as
usual, it is written as a product of a proposal probability, which
we take to be uniform among all possible final states, and an acceptance
ratio, determined by the detailed balance condition: 

\begin{equation}
\omega_{\beta\alpha}=\frac{1}{M(E_{\alpha})}\min\left(1,\frac{M(E_{\alpha})N(E_{\alpha})}{M(E_{\beta})N(E_{\beta})}\right).\label{eq:omega_alpha_beta}\end{equation}
In the master equation, the transition probabilities $\omega_{\beta\alpha}$
do not change if we vary $\beta$ (or $\alpha$) within the same energy
level. This makes it possible to write the following master equation
for the probability of having energy $E_{i}$, $P_{i}=\sum_{\alpha}p_{\alpha}\delta_{E_{\alpha},E_{i}}$\begin{eqnarray}
P_{i}(t+1) & = & \Gamma_{i}P_{i+1}(t)+\Gamma_{i-1}P_{i-1}(t)\nonumber \\
 & + & \left(1-\Gamma_{i}-\Gamma_{i-1}\right)P_{i}(t)\label{eq:master_equation}\\
 & = & \sum_{j}\Omega_{ij}P_{j}(t)\end{eqnarray}
with \begin{equation}
\Gamma_{i}=\frac{N_{i}}{M_{i+1}}\min\left(1,\frac{N_{i+1}M_{i+1}}{M_{i}N_{i}}\right).\label{eq:rates}\end{equation}
This equation defines the projection of the original Markov chain
onto to the energy variable; the random walk in energy space remains
a Markov chain with transition probabilities defined by the density
of states of the original spin model. This is,of course, an important
simplification afforded by our choice of transition rates. With the
exception of special models, like the infinite range Ising model \cite{Wu04},
the projection of conventional dynamics in energy space is non-markovian
and memory effects can have a significant impact on the dynamics \cite{DTW+04,THT04}.
Nevertheless, in the model under study, we find below that important
aspects of the long-term SSF dynamics are preserved.

Assume the system starts from an energy level $r$ at time $t=0$:\[
P_{i}(0)=\delta_{i,r}\]
and let $Q_{i}(t)$ now refer to the probability of having energy
$E_{i}$ at time $t$, given that the system has never visited energy
level $s$ ($s>r$). $Q_{i}(t)$ satisfies the master equation in
eq.~\ref{eq:master_equation} for $i<s$, with the same initial condition
as $P_{i}(t)$, namely $Q_{i}(0)=\delta_{i,r}$, and an additional
condition $Q_{s}(t)=0$ replacing eq. \ref{eq:master_equation} for
$i=s$. The probability of first passage in $s$ becomes \begin{equation}
H_{sr}(\tau)=\Gamma_{s-1}Q_{s-1}(\tau).\label{eq:tunneling_prob}\end{equation}
The master equation can be solved using a normal mode expansion, $Q_{i}(t)=\sum_{\gamma}a_{\gamma}f_{i}^{\gamma}\lambda_{\gamma}^{t}$.
The dimension of the transition matrix in energy space, $\boldsymbol{\Omega}$,
scales linearly with system size, not exponentially as in the case
of the transition matrix of the Markov process in phase space. The
diagonalization of $\boldsymbol{\Omega}$, becomes a manageable problem
allowing the calculation of the eigenvalues, $\lambda_{\gamma}$,
the left and right eigenvectors, $g_{i}^{\gamma}$and $f_{i}^{\gamma}$,
and the coefficients $a_{\gamma}=g_{r}^{\gamma}/\sum_{i}g_{i}^{\gamma}f_{i}^{\gamma}$.
It is also possible, of course, to diagonalize the transition matrix
for the actual probability distribution $P_{i}(t)$ using the same
formalism, and access the decay time of the various eigenmodes of
the master equation. The largest finite one (the equilibrium distribution,
$P_{eq}$, does not decay) is the equilibration time of $P_{i}(t)$,
$\tau_{eq}$. 

\begin{figure}
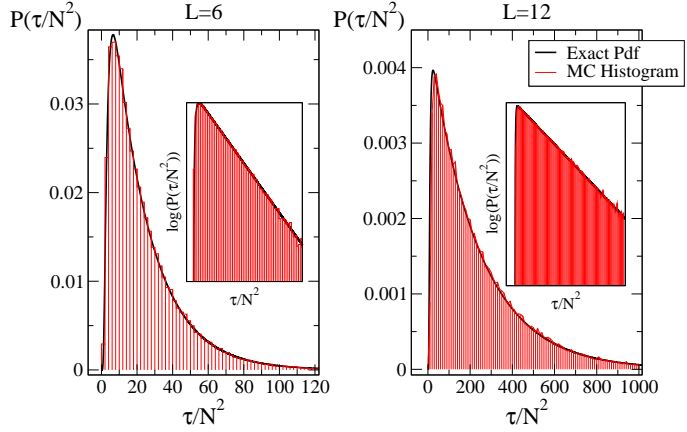

\onefigure[%
  clip,
  width=90mm]{fig1.eps}
\caption{\label{cap:convolucaoMC}Comparison between exact result given by
master equation approach and the Monte Carlo results for two spin
glass samples with $L=6$ and $L=12$. }
\end{figure}

The distribution of tunneling time measured in our simulations can
be written in the form\begin{equation}
H(\tau)=\sum_{\tau'=0}^{\tau}H_{0m}(\tau-\tau')H_{m0}(\tau')\label{eq:h_of_tau}\end{equation}
where $E_{m}$ is the energy level corresponding to the maximum of
$N(E)$. In Fig.~\ref{cap:convolucaoMC} we superpose a distribution
measured in a Monte Carlo simulation with the rates given by eq. \ref{eq:omega_alpha_beta}
with a calculated one for two spin glass samples of linear dimension
$L=6,\,12$; the simulation reproduces the calculated distribution
very accurately. 

The distribution of tunneling time between any two energy levels decays
exponentially with a time constant given by $\tau_{max}=-1/\log(\lambda_{max})$,
where $\lambda_{max}<1$ is the largest eigenvalue of the corresponding
$\boldsymbol{\Omega}$ (eq.~\ref{eq:tunneling_prob}). For a given
spin glass sample, we denote by $\tau_{u}$ and $\tau_{d}$ the longest
decay time for tunneling $E_{0}\to E_{m}$ and $E_{m}\to E_{0}$,
respectively; $H(\tau)$ (eq.~\ref{eq:h_of_tau}) will decay exponentially
with the largest of $\tau_{u}$ and $\tau_{d}$ for the given sample.
The power law decay seen in \cite{DTW+04} appears only when we aggregate
tunneling times from different realizations of the random interactions
$\pm J$. If the distribution of the largest decay time, in the ensemble
of spin glass samples, has a power law tail, $\rho\left(\tau_{max}\right)\sim\tau_{max}^{-\nu}$
for $\tau_{max}\to\infty$, we obtain, $H(\tau)\sim\int dx\rho\left(x\right)e^{-\tau/x}\sim t^{1-\nu}$
as $t\to\infty$.

\begin{figure}
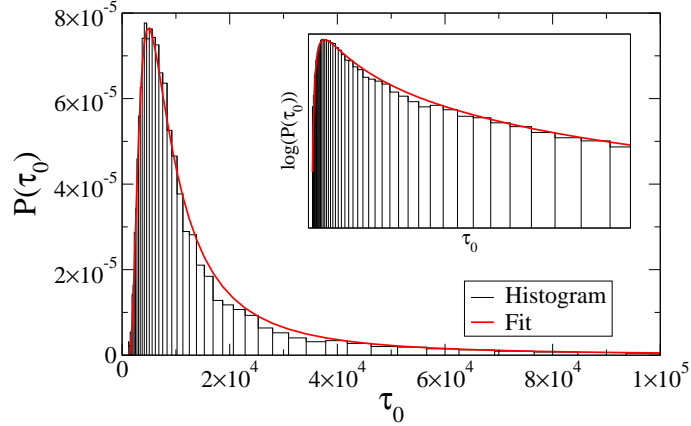

\onefigure[%
  clip,
  width=90mm,
  keepaspectratio]{fig2.eps}

\caption{\label{cap:frechet}Comparison between the histogram of $\tau_{0}$
from 10000 uniformly generated samples of $\pm J$ 2D spin glass with
$L=10$ and the maximum likelihood fit to a Frechet distribution function
obtained from the original data.}
\end{figure}

This asymptotic behaviour is directly related to features of the density
of states of the $\pm J$ spin glass. The inverse transition rate
from the ground state level to the first excited level is given by
(eq. \ref{eq:rates}),\[
\tau_{0}=\Gamma_{0}^{-1}=1+\frac{N_{1}+N_{2}}{N_{0}}.\]
It was noted in \cite{DTW+04} that $N_{1}/N_{0}$ has a Fréchet probability
distribution function, in an ensemble of spin glass samples; a similar
result occurs for $\tau_{0}$ (Fig.~\ref{cap:frechet}). This fact
completely controls the long time dynamics of the Monte Carlo simulations. 

In Fig.~\ref{cap:td_tu_teq} we plot $\tau_{eq}$, the equilibration
time, and $\tau_{u}$ and $\tau_{d}$, the decay time constants for
tunneling from $E_{0}$ to $E_{m}$ and from $E_{m}$ to $E_{0}$,
against $\tau_{0}$. In the samples with larger $\tau_{0}$, we observe
the following relations: \begin{eqnarray}
\tau_{eq} & \approx & \tau_{0}\label{eq:tau_0_tau_eq}\\
\tau_{u} & \approx & \tau_{0}.\label{eq:tau_s_tau_0}\end{eqnarray}
Note that $\tau_{d}$ can be two orders of magnitude larger than $\tau_{u}$;
tunneling from $E_{m}$ to $E_{0}$ is much slower than from $E_{0}$
to $E_{m}.$ In fact, as can be seen in the inset of Fig. \ref{cap:td_tu_teq}
(b), the following relation holds asymptotically: \begin{equation}
\tau_{d}\approx N_{b}\tau_{u}\label{eq:tau_d_tau_s}\end{equation}
where $N_{b}\sim O(N)$ is the number of energy levels between $E_{0}$
and $E_{m}$ (recall that in our simulation the energy histogram is
limited to this energy range). These results are quite simple consequences
of the existence of a bottleneck in the dynamics due to transitions
between $E_{0}$ and $E_{1}$ (large $\tau_{0}$). 

The equation for $P_{0}$ is (for long time we replace $P_{i}(t+1)-P_{i}(t)$
by $dP_{i}/dt$) \begin{equation}
\frac{dP_{0}(t)}{dt}=-\Gamma_{0}\left(P_{0}(t)-P_{1}(t)\right).\label{eq:dp_o_dt}\end{equation}
We denote the tunneling time scale between $E_{1}$ and $E_{m}$ by
$\tau_{1}$ ; for samples with very large $\tau_{0}$, we assume $\tau_{0}\gg\tau_{1}$.
For $t\gg\tau_{1}$, we will have for $i\neq0$\[
P_{i}(t)\approx\frac{1-P_{0}(t)}{N_{b}-1}=\frac{N_{b}}{N_{b}-1}P_{eq}\left(1-P_{0}(t)\right)\]
where $P_{eq}=1/N_{b}$ is the equilibrium distribution. Replacing
in eq.~\ref{eq:dp_o_dt} we obtain a time constant for the decay
of $P_{0}(t)-P_{eq}$ given by \[
\tau_{eq}=\left(\Gamma_{0}\frac{N_{b}}{N_{b}-1}\right)^{-1}\approx\Gamma_{0}^{-1}.\]

To calculate the distribution of tunneling time from $E_{0}$ to $E_{m}$,
$H(\tau)$, we use the initial condition $P_{i}(t)=\delta_{i,0}$.
Defining $P_{i}(t)=Q_{i}(t)+R_{i}(t)$, where $Q_{i}(t)$ is the probability
of having energy $E_{i}$ at time $t$ given that the system has not
reached $E_{m}$, $Q(t)=\sum_{i=0}^{N_{b-1}}Q_{i}(t)$ is the probability
that the tunneling time is greater than $t$, $Q(t)=\int_{t}^{\infty}d\tau H(\tau)$.
For $t\gg\tau_{1}$ we have\begin{eqnarray*}
P_{0}(t) & \approx & Q_{0}(t)\\
P_{i}(t) & \approx & R_{i}(t)\qquad\mbox{for }i\neq0\end{eqnarray*}
This expresses the fact that the system either has energy $E_{0}$,
and has not tunneled, or has left the ground state and has almost
certainly tunneled to $E_{m}$. Therefore\[
\frac{dQ(t)}{dt}\approx\frac{dQ_{0}}{dt}\approx\frac{dP_{0}}{dt}\]
The decay time of $Q(t)$ (and $H(t)$) is $\tau_{u}\approx\tau_{0}$.

\begin{figure}
\twofigures[%
  clip,
  width=65mm,
  keepaspectratio]{fig3.eps}{fig4.eps}

\caption{\label{cap:td_tu_teq}(a) Correlation between equilibration time
$\tau_{eq}$ and $\tau_{0}$ for lattice size $L=6$ to $20$. The
dashed line shows the asymptotic behaviour where $\tau_{eq}=\tau_{0}$.
(b) Correlation between both $\tau_{d}$ and $\tau_{u}$ with $\tau_{0}$.
The dashed line shows the asymptotic behaviour where $\tau_{u}=\tau_{0}$.
(c) Correlation between $\tau_{d}/N_{b}$ and $\tau_{u}$. The dashed
line shows the asymptotic behaviour where $\tau_{d}/N_{b}=\tau_{u}$. }

\caption{\label{cap:toy_vs_real}Correlation between $\tau_{d}/N_{b}$ and
$\tau_{u}$ obtained from Monte Carlo simulations of the $\pm J$
2D spin glass with SSF dynamics. The dashed line shows the asymptotic
behaviour where $\tau_{d}/N_{b}=\tau_{u}$. }

\end{figure}

The asymmetry of $\tau_{d}$ and $\tau_{u}$ should by now be obvious.
For tunneling from $E_{m}$ to $E_{0}$ the initial condition is $P_{i}(0)=\delta_{i,m}$
and $Q(t)=\sum_{i=1}^{m}Q_{i}(t)$. For times much larger than $\tau_{1}$
we expect, now,\[
Q_{i}(t)\approx\frac{Q(t)}{N_{b}-1}.\qquad\textrm{for }i\ne0\]
Since $dQ(t)/dt=-\Gamma_{0}Q_{1}(t)$ ($Q(t)$ only changes due to
transitions between $E_{1}$ and $E_{0}$), we obtain \[
\frac{dQ(t)}{dt}=-\frac{\Gamma_{0}}{N_{b}-1}Q_{1}(t),\]
\emph{i.e.} \[
\tau_{d}=\left(N_{b}-1\right)\tau_{u}\approx N_{b}\tau_{u}.\]

In simple language, these results can be understood as follows. Tunneling
from $E_{0}$ to $E_{m}$ is controlled by the process of exiting
the ground state level: the system cannot have tunneled if it is still
in a ground state. On the other hand, a system can only enter the
ground state level if it is in energy level $E_{1}$; therefore the
tunneling rate from $E_{m}$ to $E_{0}$ has an extra factor $P_{eq}=1/N_{b}$
corresponding to the probability of having energy $E_{1}$. 

These results provide a very clear explanation of the correlation
between tunneling time and $N_{1}/N_{0}$ found in \cite{DTW+04}.
The controlling time scale $\tau_{0}=1+\left(N_{1}+N_{2}\right)/N_{0}$
is clearly related to the ratio $N_{1}/N_{0}$. This difference in
time scales $\tau_{u}$ and $\tau_{d}$ should be present in other
models; namely, those in which the slowest processes in the simulations
involve getting across steep entropy changes. The 2D $\pm J$ spin
glass is an extreme case of this behavior, the dynamics being controlled
by the transition between the two lowest energy levels. One could
ask if this result is an artifact of our simplified dynamics. In fact,
we verified the same behavior with SSF dynamics. Fig.~\ref{cap:toy_vs_real}
shows the same type of plot as in Fig.~\ref{cap:td_tu_teq} with
averaged tunneling time (averages taken within a sample) in SSF dynamics:
for long times the relation of eq. \ref{eq:tau_d_tau_s} is still
verified. Notice that the equilibration time scale of a probability
distribution, $\tau_{eq}$, may be much smaller than tunneling time
scale, which, as usually measured, is dominated by $\tau_{d}$. This
can lead to a very pessimistic estimate of the time required to reach
equilibrium.

In summary, modelling the dynamics of a multicanonical simulation
in a sort of mean-field way, by averaging transition rates to states
of the same or nearby energies, we were able to define a Markov process
in the energy variable, reducing the dimension of the Markov matrix
to a manageable size. Nevertheless, this procedure preserves the main
features of the long time dynamics of a conventional simulation. A
very complete and physically transparent description of the more salient
features of the dynamics of multicanonical simulations becomes possible.
In particular, we clarified the relation between equilibration and
tunneling time, and the difference between tunneling \emph{away} from
regions of low density of states from tunneling \emph{into} such regions. 

\acknowledgments

The authors would like to thank E.J.S. Lage and J. Penedones for very
helpful discussions and S. Sabhapandit for providing an implementation
of Saul and Kardar's algorithm. This work was supported by FCT (Portugal)
and the European Union, through POCTI (QCA III). Two of the authors,
JVL and MDC, were supported by FCT grants numbers SFRH/BD/1261/2000
and SFRH/BD/7003/2001, respectively.

\end{document}